\documentclass[12pt,reqno]{amsart}

\usepackage{latexsym}
\usepackage{graphicx}

\newcommand{\R}{{\mathbb R}}

\newcommand{\Z}{{\mathbb Z}}

\renewcommand{\phi}{\varphi}

\newcommand{\sgn}{\mathop{{\rm sgn}}\nolimits} 
\newcommand{\sg}{\frak{S}}  
\newcommand{\ol}{\overline}  
\newcommand{\tr}{{\rm tr}} 
\newcommand{\go}{\mathfrak}

\newcommand{\ep}{\varepsilon}

\newtheorem{theo}{{\sc Theorem}}[section]

\newtheorem{conj}[theo]{{\sc Conjecture}}

\newtheorem{lem}[theo]{{\sc Lemma}}
\newtheorem{prop}[theo]{{\sc Proposition}}

\title[Counter-example to  conjectured   $SU(N)$ character asymptotics]
{Counterexample to  conjectured  $SU(N)$ character asymptotics}

\author{Tatsuya Tate}
\address{Department of Mathematics, Keio University, Keio University
3-14-1 Hiyoshi Kohoku-ku, Yokohama, 223--8522 Japan}
\email{tate@math.keio.ac.jp}

\author{Steve Zelditch }
\address{Department of Mathematics, Johns Hopkins University, Baltimore, MD
21218, USA}
\email{zelditch@math.jhu.edu}

\thanks{Research partially supported by JSPS (first author).}
\thanks{Research partially supported by NSF grants DMS-0071358 and DMS-0302518 and by the Clay Foundation (second author).}

\date{\today}

\begin{document}

\maketitle

\section{Introduction} The purpose of this note is to give a counterexample
to the  conjectured large $N$ asymptotics of character values
$\chi_R(U)$ of irreducible characters of $SU(N)$, which appears in
papers  of Gross-Matytsin \cite{M, GM} and Kazakov-Wynter
\cite{KW}. Asymptotics of characters are important in the large
$N$ limit in  $YM_2$ ($2D$ Yang-Mills theory) and in certain
matrix models \cite{KSW, KSW2, KSW3}). Our counterexample consists
of one special sequence of elements $a_N \in SU(N)$ for which the
conjectured asymptotics on $\chi_{R_N}(a_N)$ fail for any relevant
sequence $\chi_{R_N}$ of irreducible characters.  It is not clear
at present how widespread in $SU(N)$ the failure is.

To state the conjecture and the counterexample, we will need some
notation.  We recall that irreducibles  of $SU(N)$ are
parametrized by their highest weights $\lambda$ or equivalently by
Young diagrams with $\leq N - 1$ rows. To facilitate comparison
with  \cite{GM}, we will use a further parametrization of
representations $R$ of $SU(N)$ by their shifted highest weights
\begin{equation} \label{ell} \ell = \lambda + \rho_N,\;\;\; \rho_N = \;\; \mbox{half the
sum of the positive roots}. \end{equation}  The components of the
shifted highest weight are then strictly decreasing  $ \infty
> \ell_1 > \ell_2 > \cdots > \ell_N > -\infty  $ (cf. \eqref{hs} for the explicit formula).
 To a shifted highest weight we associate the probability
measure on $\R$ defined by \begin{equation} \label{DRHOR} d \rho_R
= \frac{1}{N} \sum_{j = 1}^N \delta(\frac{\ell_j}{N}),\ i.e.
\;\;\int_{\R} f(y) d\rho_R(y) = \frac{1}{N} \sum_{j = 1}^N f
(\frac{\ell_j}{N}). \end{equation}  Given a sequence $R_N$ of
irreducible representations of $SU(N)$, we write \begin{equation}
\label{TENDS} R_N \to d \rho, \;\;\mbox{ if }\;\;  d\rho_R \to
d\rho\;\; \mbox{ in the sense of measures}. \end{equation} Any
weak limit is a probability measure satisfying $\rho_Y([0, T])
\leq T$, since $\ell_{j} - \ell_{j+1} \geq 1.$ If the limit has a
density, which is written $d \rho_Y = \rho_Y' (y) dy$, then
$\rho_Y' (y) \leq 1.$ A limit measure is called a
 ``distribution  on Young tableaux".

The conjecture of Gross-Matytsin, Kazakov-Wynter and other
physicists   concerns the values of a sequence of characters
$\chi_{R_N}$ on elements $U_N$ of $SU(N)$.
 The
eigenvalue distribution of $U \in SU(N)$ with eigenvalues $\{e^{i
\theta_k}\}$  is the probability measure on the unit circle $S^1$
defined
$$d\sigma_N := \frac{1}{N} \sum_{k = 1}^{N} \delta (e^{i
\theta_k}). $$ Given a sequence $U_N \in SU(N)$, we write  $U_N
\to \sigma$ (as $N \to \infty)$ if $d\sigma_N  \to
 \sigma
 $
 in the sense of measures, i.e. $\frac{1}{N} \sum_{k = 1}^{N} f (e^{i
\theta_k}) \to \int_{S^1} f d
 \sigma.
 $

\begin{conj}\label{GM} (Gross-Matytsin \cite{GM}, (2.3); Kazakov-Wynter \cite{KW}, Appendix 5.1;
\cite{KSW}, \S 3;  see below) Assume $U_N \to \sigma, R_N \to
\rho$. Then $\chi_{R_N}(U_N) \sim e^{N^2 F_0[\rho,  \sigma)]}$
where
$$\begin{array}{l} F_0(\rho, \sigma) = S(\rho, \sigma) + \frac{1}{2}
\{\int_{\R} \rho(x) x^2 dx + \int \sigma(y) y^2 dy\} \\ \\
- \frac{1}{2} \{\int_{\R \times \R} \rho(x) \rho(y) \ln |x - y| dx
dy  + \int_{\R \times \R} \sigma(x) \sigma(y) \ln |x - y| dx dy\},
\end{array}$$ where $S$ is the classical action corresponding to
the Hopf equation
$$\left\{ \begin{array}{l} \frac{\partial f}{\partial t} + f \frac{\partial f}{\partial x} = 0\\ \\
\Im f(x, 0) = \pi \rho(x),\;\; \Im f(x, 1) = \pi \sigma(x).
\end{array} \right. $$

\end{conj}

Our counterexample is based on the  special sequence  $U_N = a_N$
of {\it principal elements of type $\rho$} of $SU(N)$ in the sense
of Kostant \cite{Ko}. Such a principal  element is regular and has
minimal order in $SU(N)$, given by its Coxeter number $N$.  The
eigenvalues of $a_N$ are thus the distinct $N$th roots of unity,
and the  limit distribution $d\sigma$ of $a_N$  of eigenvalues is
obviously $d \theta$.

 The key fact,
discovered by Kostant \cite{Ko} (see also \cite{Ko2, AF} is that
characters take on only the three values
\begin{equation}\label{ID}  \chi_R(a_N) = 0, \pm 1,\;\;\; \forall R \in
\widehat{U(N)}. \end{equation} This immediately casts doubt on the
conjecture, since it would imply:
\begin{equation} e^{N^2 F_0[\rho,  d\theta)]} \sim \chi_{R}(a_N) = \left\{ \begin{array}{ll} (i) & 0\\& \\
(ii) & -1\\& \\
(iii) & 1  \end{array} \right.,\;\;\; \forall \rho. \end{equation}
Clearly, this would require that, for all $d\rho$,
\begin{equation} F_0[\rho,  d\theta)] \; \mbox{is} \; \left\{ \begin{array}{ll}(i)  & < 0 \\ &\\
(ii) & = i \pi (2k_N + 1)\\ & \\(iii) & = o(1/N^2)  \end{array}
\right.
\end{equation}

The following result shows that the oscillation of values of
$\chi_R(a_N)$ is much too regular for any such results. There is
simply no separation of the possible limiting shapes of Young
diagrams into the three discrete classes of possible limits $0,
\pm 1$; all possible limit shapes are consistent with the limit
$0$.

\begin{theo}
\label{main1} Given any sequence  of irreducibles  $R_N \in
\widehat{SU(N)}$, with $R_N \to \rho$, there exists a sequence
$R_N' \in \widehat{SU(N)}$ with $R_N' \to \rho$ with the property
that $\chi_{R_N'}(a_N) = 0$. Hence, there cannot exist a limit
functional $F_0(d\theta, d\rho)$ depending only on the limit
densities $d\sigma, d\rho$.
\end{theo}

The basic idea of the proof is the following: suppose that the
highest weight  $R$ is such that   $\chi_R(a_N) = \pm 1$. Then, by
changing one component of $R$ by one unit, one obtains a highest
weight $R'$  such that  $\chi_{R'}(a_N) = 0$. Taking a sequence
$R_N \to \rho$ and changing $R_N \to R_N'$ one obtains a new
sequence with $R_N' \to \rho$ and with $\chi_{R_N'}(a_N) \equiv
0.$

\subsection{Background of the conjecture}

The Conjecture \ref{GM} attributed above  to Gross-Matytsin and
Kazakov-Wynter seems to have appeared independently  in the papers
\cite{GM} and \cite{KW}. It is analogous to and inspired by
Matytsin's conjecture \cite{M} on the large $N$ asymptotics of
Itzykson-Zuber integrals. The latter conjecture has recently been
proved by Guionnet-Zeitouni \cite{GZ} and Guionnet \cite{G}. But
the former is incorrect in general.  We now explain the difference
between the two conjectures and give some background on the
context in which the conjecture arose.

  The original conjecture of Matytsin
pertained to integrals known variously as Itzykson-Zuber or
spherical integrals \begin{equation} \label{IZ} I(A, B)\equiv
\int_{SU(N)} e^{N\tr[A U B U^{\dagger}]} dU,
\end{equation} where $A$ and $B$ are $N \times N$ Hermitian
matrices  and $dU$ is (unit mass)
 Haar measure on $SU(N)$. By the Itzykson -- Zuber (Harish-Chandra) formula one
has
\begin{equation} I(A, B) ={{{\rm det}[e^{Na_{i}
b_{j}}]}\over{\Delta(a)\Delta(b)}}, \end{equation} where $\{a_i\},
$ resp. $\{b_j\}$, are the eigenvalues of $A$, resp. $B$ and where
$\Delta(a)$ denotes the Van der Monde determinant
$\Delta(a)=\Pi_{i<j}(a_{i}-a_{j})$.  In \cite{M}, Matysin stated
Conjecture \ref{GM} precisely in the same form for $I_N(A_N,
B_N)$. This conjecture has recently been proved by
Guionnet-Zeitouni \cite{GZ} and Guionnet \cite{G}.

In the subsequent papers \cite{GM, KW}, Gross-Matytsin and
Kazakov-Wynter  stated an analogous conjecture for characters of
$U(N)$. We quote their statements in some detail to draw attention
to the key difference to Matytsin's original conjecture.

First, we consider \cite{GM}. There are some slight differences in
notation (e.g. their $\Xi$ is our $F_0$) which we leave to the
reader to adjust. They write: ``for large $N$ the $U(N)$
characters behave asymptotically as
\begin{equation}  \chi_R(U)\simeq{\rm e}^{N^2 \Xi[\rho_Y(l/N),
\sigma(\theta)] }\end{equation} with some finite functional
$\Xi[\rho_Y, \sigma]$. In this formula it is implicit that we take
the limit $N\to\infty$ assuming that the eigenvalue distribution
of the unitary $N\times N$ matrix $U$ converges to a smooth
function
 $\sigma(\theta)$, $\theta\in [0, 2\pi]$. (The eigenvalues of a unitary matrix lie on the  unit
circle in the complex plane and can be parametrized as
$\lambda_j={\rm e}^{i\theta_j}$.) In addition, it is assumed that
the distribution of parameters ${\tilde y}_i=l_i/N$, which define
the representation  $R$, also converges to another smooth function
$\rho_Y({\tilde y})$, that we can call the Young tableau  density.
The functional $\Xi$ is, in general, not easy to calculate.
However, in some important cases it can be found explicitly." They
continue: ``...we will have to evaluate the functional derivatives
of $\Xi[\rho_Y, \sigma_1]$... This can be done if we observe that
the $U(N)$ characters  can be represented as analytic
continuations of the Itzykson--Zuber integral (\ref{IZ}). Setting
$a_k=l_k$, $b_j=\theta_j$ and analytically continuing $a_k\to i
a_k$, we see that \begin{equation}\label{KCF} {{{\rm
det}[e^{Na_{i} b_{j}}]}\over{\Delta(a)\Delta(b)}} \rightarrow
J\big({\rm e}^{i\theta_s}\big)\, \chi_R(U).
\end{equation}

Therefore, we can use  the known expressions for the large $N$
limit of the Itzykson--Zuber integral  to find the functional
$\Xi$. In particular, if  as $N\to\infty$ the distributions of
$\{a_k\}$ and $\{b_j\}$ converge to smooth functions $\alpha(a)$
and $\beta(b)$, then asymptotically..." the formula in Theorem
\ref{main1} holds with $\alpha da = \rho, \beta db = \sigma.$

In \cite{KW}, it is pointed out that character values for $U(N)$
are, `up to a factor of $i$...the Itzykson-Zuber
determinant...From Matysin's paper we quote the result (with the
minor change of an extra factor of $i$)...'

We note that the relation (\ref{KCF}) between Itzykson-Zuber
integrals and characters is the Kirillov character formula, see
e.g. Theorem 8.4 of \cite{BGV}.  Thus, it is  precisely the
analytic continuation of the large $N$ asymptotics of the Itzykson
-Zuber integral from Hermitian to skew-Hermitian matrices (the Lie
algebra of $U(N)$), i.e. the extra factor of $i$,  which leads in
general to incorrect results. The same error then propagates to
the conjecture of Gross-Matytsin and Kazakov-Wynter on the large
$N$ asymptotics of the partition function of  $2$D $SU(N)$
Yang-Mills theory on a cylinder, which the second author disproved
by a related counterexample in \cite{Z}. On the positive side, the
proof of Guionnet-Zeitouni of Matytsin's conjecture suggests that
the conjectured partition function asymptotics might be correct
after analytically continuing the partition function from $U(N) $
to positive matrices.

In the study of matrix models, Kazakov-Staudacher-Wynter
\cite{KSW, KSW2, KSW3} employ related asymptotics related to
 for matrices satisfying  certain moment conditions
(i.e. on traces of powers). These do not appear to exclude unitary
matrices. It is not clear if  they exclude the counterexamples we
are presenting.

Of course, the  counterexample does not indicate the limit of
validity of the original conjectures or of their applications in
$2D$ gravity, $YM_2$ and  matrix models. V.  Kazakov has raised a
number of interesting questions regarding the counterexample. Can
one perturb the counterexample or does it depend on the
eigenvalues being roots of unity?
 Are the conjectures even  `generically correct'  in a reasonable sense?
 Rather than studying pointwise limits, one can study asymptotics
 of statistical aspects of character values. The
large deviations theory of Guionnet-Zeitouni \cite{GZ, G} does not
seem to  adapt in a straightforward way to character values on
$SU(N)$. What is a good probabalistic framework?   Some
interesting work in the statistical  direction is found in the
works of Kazakov-Staudacher-Wynter (loc. cit.). M. R. Douglas has
suggested a different point of view, connecting large $N$ limits
with conformal field theory \cite{D1, D2}.

\vspace{5pt}

\noindent {\bf Acknowledgment} This note was begun during a stay
of the first author as a JSPS fellowship at the Johns Hopkins
University and was continued while both authors were visiting MSRI
as part of the Semiclassical Analysis program. The second author
was partially supported by the Clay Foundation. We   thank M.
Douglas and particularly  V. Kazakov for  many  comments and
questions.

\section{Review of the Kostant identity}

In this section, we shall review the Kostant identity (\ref{ID})
for the values $\chi_R(a_N)$ of irreducible characters at the
principal elements of type $\rho$.  We  also review a version of
the formula obtained in \cite{AF} for the group $SU(N)$.

\subsection{The Kostant identity in general}

Let $G$ be a compact, connected, simply-connected semisimple Lie
group. We also assume that $G$ is simply-laced, that is, each root
has the same length with respect to the Killing inner product. Let
$a_{\rho}$ denote a principal element of type $\rho$. The element
$a_{\rho}$ is in the conjugacy class of the element
$\exp(\kappa^{-1}(2\rho))$, where $\rho$ is half the sum of the
positive roots and $\kappa$ is the isomorphism between the Lie
algebra of a fixed maximal torus and its dual induced by the
Killing form. Since the characters are  class functions,  we can
set $a_{\rho}=\exp (\kappa^{-1}(2\rho))$.

Let $\Lambda^{*}$ denote the root lattice. Let $h$ be the Coxeter
number. The number $h$ is defined as the order of the Coxeter
element in the Weyl group $W$, namely the element
$s_{\alpha_{1}}\cdots s_{\alpha_{l}}$, where $\{\alpha_{j}\}$ are
the  simple roots, $s_{\alpha_{j}} \in W$ is the  reflection
corresponding to $\alpha_{j}$ and $l$ is the rank of $G$. The
following lemma (Lemma 3.5.2 in \cite{Ko}) is one of the key
points of  \cite{Ko}.

\begin{lem}
\label{KOlem}
Let $\lambda$ be a dominant weight.
Then, either
\begin{enumerate}
\item[{\rm (1)}] For all $w \in W$, $w(\lambda +\rho)-\rho \not \in h\Lambda^{*}$ or
\item[{\rm (2)}] There exists a unique $w \in W$ such that $w(\lambda +\rho)-\rho \in h\Lambda^{*}$.
\end{enumerate}
\end{lem}

It should be noted that, if $\lambda$ satisfies the condition $(2)$ in Lemma \ref{KOlem},
then $\lambda \in \Lambda^{*}$, since $w\rho-\rho$ is in the root lattice $\Lambda^{*}$ for all $w\in W$,
and the lattice $h\Lambda^{*}$ is invariant under $W$-action.

By using Lemma \ref{KOlem}, we define $\ep(\lambda) \in \{0,\pm 1\}$ for each dominant weight $\lambda$ as follows:
\begin{equation}
\label{EPS}
\ep(\lambda)=
\left\{
\begin{array}{ll}
\sgn(w) & \mbox{if $\lambda$ satisfies $(2)$ in Lemma \ref{KOlem}},\\
0 & \mbox{otherwise}.
\end{array}
\right.
\end{equation}

Then, the Kostant identity can be stated as follows:

\begin{theo}[Kostant\cite{Ko}]
\label{KOth} Under the assumption on $G$ stated above, the
irreducible characters $\chi_{\lambda}$ take one of the values
$0$, $1$ or $-1$ at the element $a_{\rho}$. More precisely, one
has
\[
\chi_{\lambda}(a_{\rho})=\ep(\lambda)
\]
for each dominant weight $\lambda$.
\end{theo}

\subsection{The Kostant identity for $SU(N)$}
Now we set $G=SU(N)$. In this case, a dominant weight is regarded as a partition of a non-negative integer,
and one can rewrite Theorem \ref{KOth} in terms of a property of components of partitions.
We refer the readers to \cite{FH} for a general theory of the representation theory
of $SU(N)$ and partitions, and to \cite{AF} for a version of Kostant's theorem (Theorem \ref{KOth})
for $SU(N)$, which we shall review in this section.

To fix notation, we first define a correspondence between the
dominant weights and  partitions. Let $\go{t}_{N}$ and
$\go{h}_{N}$ denote the Lie algebras of maximal tori in $SU(N)$
and $U(N)$ respectively, and let $L_{N}^{*}$ and $I_{N}^{*}$
denote the weight lattices in the dual spaces $\go{t}_{N}^{*}$ and
$\go{h}_{N}^{*}$ respectively. We denote the standard basis in
$\go{h}_{N}^{*}$ by $e_{j}$, $j=1, \ldots, N$. Then, for each $\mu
\in L_{N}^{*}$, there is a unique $f=f_{\mu} \in I_{N}^{*}$ such
that
\[
f=\sum_{j=1}^{N-1}f_{j}e_{j},\quad f|_{\go{t}_{N}}=\mu.
\]
Therefore, the weight lattice $L_{N}^{*}$ is identified with the sublattice (of rank $N-1$) in $I_{N}^{*}$
spanned by $e_{1},\ldots,e_{N-1}$:
\begin{equation}
\label{WL}
L_{N}^{*} \cong \bigoplus_{j=1}^{N-1}\Z \cdot e_{j} \subset I_{N}^{*}.
\end{equation}

The roots for $(\go{su}(N),\go{t}_{N})$ are given by the
restrictions to $\go{t}_{N}$ of the following elements in
$I_{N}^{*}$ (which are the roots for $(\go{u}(N),\go{h}_{N})$):
$\alpha_{i,j}=e_{i}-e_{j},\quad 1 \leq i \neq j \leq N$. We take
the  positive roots to be  $\alpha_{i,j}$ ($i<j$), and  the simple
roots to be  $\alpha_{j}:=\alpha_{j,j+1}$, $j=1,\ldots,N-1$. The
corresponding  positive open Weyl chamber $C$ is given, in terms
of the identification \eqref{WL}, by
\[
C=\{f=\sum_{j=1}^{N-1}f_{j}e_{j} \in \go{t}_{N}^{*}\,;\,f_{1} > \cdots > f_{N-1} > 0\}.
\]
Thus, the set of dominant weights $P_{N}:=\ol{C} \cap L_{N}^{*}$ is given by
\[
P_{N}=\{f=\sum_{j=1}^{N-1}f_{j}e_{j}\,;\,f_{1} \geq \cdots \geq f_{N-1} \geq 0,\quad f_{j} \in \Z\},
\]
which is the set of partitions of length $N$ whose last component is zero.
In this notation, half the sum of the positive roots $\rho_{N}$ is given by
\begin{equation}
\label{hs}
\rho_{N}=\sum_{j=1}^{N-1}(N-j)e_{j}.
\end{equation}
The principal element of type $\rho$, $a_{N}:=\exp (\kappa^{-1}(2\rho_{N}))$,
is given by
\begin{equation}
a_{N}={\rm diag}(e^{\pi i(N-1)/N},e^{\pi i(N-3)/N},\ldots,e^{-\pi i(N-3)/N},e^{-\pi i(N-1)/N}),
\label{pe}
\end{equation}
and, in particular, the distribution of the eigenvalues of $a_{N}$
tends to the normalized Haar measure on the circle.

The Kostant identity (Theorem \ref{KOth}) can be rewritten in the following form,
which is obtained in \cite{AF}:

\begin{prop}[\cite{AF}]
\label{KOSU}
Let $\lambda$ be a dominant weight for $SU(N)$,
and let $\chi_{\lambda}$ be the irreducible character for $SU(N)$ corresponding to $\lambda$.
As above, we write $\lambda=(\lambda_{1},\ldots,\lambda_{N-1},\lambda_{N})$ with
$\lambda_{N}=0$.
Let $a_{N}=\exp(\kappa^{-1}(2\rho_{N}))$. Then, $\chi_{\lambda}(a_{N}) \neq 0$ if and only if
$\lambda_{j} +N-j$'s have distinct residue modulo $N$. In such a case, we have
\[
\chi_{\lambda}(a_{N})=\sgn(\sigma),
\]
where $\sigma \in \sg_{N}$ is defined by
\[
\sigma(j)=N-r(j),\quad j=1,\ldots,N,
\]
and $r(j)$ denotes the residue of $\lambda_{j}+N-j-\ell$ modulo $N$ with $\ell=|\lambda|/N$.
\end{prop}

Note that if $\lambda_{j}+N-j$'s have distinct residue modulo $N$, then
$|\lambda|$ is automatically a multiple of $N$.
In Proposition \ref{KOSU}, the numbers $\lambda_{j}+N-j$ are the components
of the dominant weight $\lambda +\rho_{N}$:
\[
\lambda +\rho_{N}=\sum_{j=1}^{N-1}(\lambda_{j}+N-j)e_{j}.
\]
The shifted highest weight $\lambda +\rho_{N}$ is writen as $\ell$ in (\ref{ell}).

\section{Proof of Proposition \ref{KOSU} and Theorem \ref{main1} }

We prepare for the proofs with a series of Lemmas. For $SU(N)$, it
is easy to see that the Coxeter number $h$ is equal to $N$. This
is proved, for example, by showing that the Coxeter element is
just a cycle of length $N$.

The following condition specializes  condition $(2)$ in Lemma
\ref{KOlem} to $SU(N)$:
\[
\mbox{$(K)$\ \ \ \ \
there exists a unique $w \in \sg_{N}$ such that $w(\lambda +\rho_{N})-\rho_{N} \in N\Lambda^{*}$.}
\]

\begin{lem}
\label{ROOTL}
Let $\mu \in L_{N}^{*}$. Then $\mu \in \Lambda^{*}$ if and only if $|f_{\mu}| \in N\Z$,
where $|f_{\mu}|=\sum_{j=1}^{N-1}f_{j}$, $f_{\mu}=\sum_{j=1}^{N-1}f_{j}e_{j} \in I_{N}^{*}$,
$f_{\mu}|_{\go{t}_{N}}=\mu$.
\end{lem}

\begin{proof}
We set $e_{0}=\sum_{j=1}^{N}e_{j}$ which is a weight for $\go{u}(N)$,
and also set $H_{0}=\sum_{j=1}^{N}H_{j}$, where $H_{j}$ is the standard basis for
the Lie algebra $\go{h}_{N}$ of the maximal torus in $U(N)$.
Then, we have $\go{t}_{N}^{*}=\go{h}_{N}^{*}/\R e_{0}$. We first claim that
\begin{equation}
\label{aux222}
\Lambda^{*}=\{f \in I_{N}^{*}\,;\,f(H_{0})=0\}/\Z e_{0}.
\end{equation}
To prove \eqref{aux222}, we recall that the root lattice $\Lambda^{*}$ is spanned by
the simple roots:
\begin{equation}
\label{aux444}
\Lambda^{*}=\bigoplus_{j=1}^{N-1}\Z \alpha_{j}, \quad
\alpha_{j}=e_{j}-e_{j+1}.
\end{equation}
Thus, any $\mu \in \Lambda^{*}$ is expressed as
$\mu=\sum_{j=1}^{N-1}c_{j}\alpha_{j}$, $c_{j} \in \Z.$
We define $f_{\mu} \in I_{N}^{*}$ by
\[
f_{\mu}=c_{1}e_{1} +\sum_{j=2}^{N-1}(c_{j}-c_{j-1})e_{j} +c_{N-1}e_{N}.
\]
Then, clearly we have $f_{\mu}(H_{0})=0$ and $f_{\mu}|_{\go{t}_{N}}=\mu$, which shows \eqref{aux222}.

Now, let $\mu \in L_{N}^{*}$. As before, we identify $\mu$ with a weight $f =f_{\mu} \in I_{N}^{*}$ of the form:
\[
f=\sum_{j=1}^{N-1}f_{j}e_{j},\quad f_{j} \in \Z,\quad j=1,\ldots,N-1,\quad
f|_{\go{t}_{N}}=\mu.
\]
In the above, we sometimes set $f_{N}=0$.

First, assume that $|f|=f(H_{0}) \in N\Z$.
We define a weight $g=g_{f}=\sum_{j=1}^{N}g_{j}e_{j} \in I_{N}^{*}$ by
\begin{equation}
\label{aux111}
g_{N}=-\frac{1}{N}\sum_{j=1}^{N-1}f_{j},\quad
g_{j}=f_{j}+g_{N},\quad j=1,\ldots,N-1.
\end{equation}
Then, by the assumption that $|f| \in N\Z$, $g_{j}$ is an integer for every $j=1,\ldots,N$.
It is easy to see that $\sum_{j=1}^{N}g_{j}=g(H_{0})=0$ and $g|_{\go{t}_{N}}=f|_{\go{t}_{N}}=\mu$.
Thus, by \eqref{aux222}, we have $\mu \in \Lambda^{*}$.
Conversely, assume that $\mu \in \Lambda^{*}$.
Then, by \eqref{aux222}, there exists a $g=\sum_{j=1}^{N}g_{j}e_{j} \in I_{N}^{*}$ such that
$g(H_{0})=0$ and $g|_{\go{t}_{N}}=\mu$.
We define $f=f_{g}=\sum_{j=1}^{N-1}f_{j}e_{j}$ by $f_{j}=g_{j}-g_{N}$ for $j=1,\ldots,N-1$.
Then, clearly, $f|_{\go{t}_{N}}=g|_{\go{t}_{N}}=\mu$, and we have
$|f|=f(H_{0})=-Ng_{N} \in N\Z$, which completes the proof.
\end{proof}

The following Lemma \ref{Clat} can be shown easily by using Lemma \ref{ROOTL}.

\begin{lem}
\label{Clat}
Let $\mu \in L_{N}^{*}$, and let $f=f_{\mu}=\sum_{j=1}^{N-1}f_{j}e_{j}$ be the corresponding weight in
$I_{N}^{*}$. Then, $\mu \in N\Lambda^{*}$ if and only if $f_{j} \in N\Z$ and $|f| \in N^{2}\Z$.
\end{lem}

\begin{lem}
\label{CL}
Let $\lambda \in P_{N}$ be a dominant weight.
We denote, as before, by $|\lambda|$ the sum $\sum_{j=1}^{N-1}\lambda_{j}$ for the representative
$f=f_{\lambda}=\sum_{j=1}^{N-1}\lambda_{j}e_{j}$ of $\lambda$.
Assume that $|\lambda|=N\ell$ with a non-negative integer $\ell$
(so that, by Lemma \ref{ROOTL}, $\lambda \in \Lambda^{*}$).
Then, the dominant weight $\lambda$ satisfies the condition $(K)$ if and only if
there exists a permutation $w \in \sg_{N}$ such that
\[
\lambda_{w(j)}+j-w(j)-\ell \in N\Z,\quad j=1,\ldots,N,
\]
where we set, as before, $\lambda_{N}=0$.
In the above condition, we can take the same permutation $w$ as that in the condition $(K)$.
\end{lem}

\begin{proof}
First, assume that $\lambda$ satisfies the condition $(K)$, and let $w \in \sg_{N}$ denote the permutation
in the condition $(K)$. We set $\mu=w(\lambda +\rho_{N})-\rho_{N} \in N\Lambda^{*}$.
Then, one has
\[
\mu=\sum_{j=1}^{N}[\lambda_{w(j)}+j-w(j)]e_{j}.
\]
(Strictly speaking, the above expresses one of the representative of $\mu \in N\Lambda^{*}$.)
We express the above weight in $I_{N}^{*}$ as an element in ${\rm span}_{\Z}(e_{1},\ldots,e_{N-1})$.
We have $\mu=\sum_{j=1}^{N-1}\mu_{j}e_{j}$ with
\begin{equation}
\label{aux5}
\mu_{j}=\lambda_{w(j)}+j-w(j) -(\lambda_{w(N)}+N-w(N)).
\end{equation}
Since $\mu \in N\Lambda^{*}$, by Lemma \ref{Clat}, we have $\mu_{j} \in N\Z$ and $|\mu| \in N^{2}\Z$.
By \eqref{aux5}, we have
\begin{equation}
\label{aux6}
|\mu|=|\lambda|-N\lambda_{w(N)}+Nw(N)-N^{2}=-N(\lambda_{w(N)}+N-w(N)-\ell),
\end{equation}
which shows that $\lambda_{w(N)}+N-w(N)-\ell \in N\Z$. Again by \eqref{aux5}, we have
\[
\mu_{j} \equiv \lambda_{w(j)}+j-w(j) \equiv 0 \mod N.
\]
Conversely, assume that there exists a $w \in \sg_{N}$ satisfying the condition in the lemma.
Then, one can write
\[
\lambda_{w(j)} +j-w(j) -\ell =Nc_{j},\quad c_{j} \in \Z, \quad j=1,\ldots,N.
\]
We set $\mu=w(\lambda +\rho_{N})-\rho_{N}$. Then, \eqref{aux5} and \eqref{aux6} still hold for this $\mu$,
and which show that $\mu_{j} \in N\Z$ and $\sum \mu_{j} \in N^{2}\Z$.
\end{proof}

\subsection{Proof of Proposition \ref{KOSU}} First of
all, we assume that the dominant weight $\lambda$ satisfies the
condition $(K)$. By Lemma \ref{ROOTL} and the fact that
$N\Lambda^{*} \subset \Lambda^{*}$, we have $|\lambda| \in N\Z$.
We set $\ell =|\lambda|/N \in \Z$. Then, by Lemma \ref{CL}, the
permutation $w \in \sg_{N}$ in the condition $(K)$ satisfies
$\lambda_{j}+w^{-1}(j)-j-\ell \in N\Z$ for any $j=1,\ldots,N$,
where we have replaced $j$ by $w^{-1}(j)$ in the statement of
Lemma \ref{CL}. We write
\[
\lambda_{j}+N-j-\ell =Na_{j}+N-w^{-1}(j).
\]
Since $0 \leq N-w^{-1}(j) \leq N-1$ are all distinct, the above equation shows that $N-w^{-1}(j)$
is the residue of $\lambda_{j}+N-j-\ell$ modulo $N$, and the residues are distinct.
Thus the residues of $\lambda_{j}+N-j$'s are also distinct.
Conversely, assume that the residues of $\lambda_{j}+N-j$'s modulo $N$ are distinct.
Denote their residues modulo $N$ by $c_{j}$, $0 \leq c_{j} \leq N-1$, $j=1,\ldots,N$.
Then, one has
\[
|\lambda| +\frac{N(N-1)}{2} \equiv
\sum_{j=1}^{N} c_{j} \equiv \frac{N(N-1)}{2} \mod N,
\]
which shows that $\ell :=|\lambda|/N$ is a non-negative integer.
We denote by $r(j)$, $0 \leq r(j) \leq N-1$ the residue of $\lambda_{j}+N-j-\ell$ modulo $N$.
We define the permutation $w \in \sg_{N}$ by
\[
w^{-1}(j)=N-r(j),\quad j=1,\ldots,N.
\]
Now, it is easy to see that $\lambda_{j}+w^{-1}(j)-j-\ell \in N\Z$, and hence, by Lemma \ref{CL},
$\lambda$ satisfies the condition $(K)$.
By Theorem \ref{KOth}, we have
\[
\chi_{\lambda}(a_{N})=\sgn(w^{-1})=\sgn(\sigma),
\]
where $\sigma=w^{-1}$ is defined in Proposition \ref{KOSU}.
This completes the proof.
\hfill\qedsymbol

\vspace{10pt}

\subsection{ Proof of Theorem  \ref{main1}}

 This is a direct consequence of Proposition
\ref{KOSU}. In fact, let $\lambda(N)$ be a sequence of dominant
weights such that $\lambda(N) +\rho_{N}$ tends weakly to a measure
$\rho_{Y}$ on the real line in the sense of (\ref{TENDS}).
Note that we have
$\lambda_{1}(N) \geq \lambda_{2}(N)$, and hence the weight $\mu(N)
=  \lambda(N) +e_{1}$ is also a dominant weight. Then we need to show that
\begin{itemize}

\item (i)
the sequence of shifted  dominant weights $\mu(N) + \rho_N
=\lambda(N)+ e_1 + \rho_{N}$ converges weakly to the same density
$\rho_{Y}$ as for the sequence $\lambda(N) + \rho_N$,  and that

\item (ii) $\chi_{\mu(N)}(a_{N})=0$.

\end{itemize}

To prove (ii), we observe that the   residues of two components
modulo $N$ of the dominant weight $\mu(N) + \rho_N$ must coincide
because the the residues of the components of $\lambda(N)+\rho_{N}$
are all distinct, and the residues of the components of $\mu(N) +\rho_{N}$
differ from that of $\lambda(N) +\rho_{N}$ only in the first component. Thus,
by Proposition \ref{KOSU}, we have $\chi_{\mu(N)}(a_{N})=0$.

To prove (i), we let $f$ be a compactly supported continuous
function on the real line, and we denote by $d \rho_{\mu_N}, $
resp. $d \rho_{\lambda_N}$ the measures in (\ref{DRHOR}) for the
corresponding irreducibles.  Then we clearly have
\[ |\int_{\R} f(x) [d \rho_{\mu_N} - d \rho_{\lambda_N}]| =
\frac{1}{N} |f(\lambda_{1}(N)/N+1)-f(\lambda_{1}(N)/N+1-1/N)| \to
0, \quad N \to \infty.
\]
Hence the sequence of the dominant weights $\mu(N)+\rho_{N}$ tends
to the same limit as the limit of the sequence
$\lambda(N)+\rho_{N}$. \hfill\qedsymbol

\end{document}